\documentclass[showpacs,twocolumn,pre,floatfix]{revtex4}
\usepackage{psfrag,epsfig,amsfonts,amssymb,amsmath,wasysym,bm}
\usepackage{dcolumn}

\usepackage[normalem]{ulem}
\usepackage{color}

\newcommand{\lsim}{\mathrel{\hbox{\rlap{\lower.55ex \hbox{$\sim$}} \kern-.3em \raise.4ex \hbox{$<$}}}}
\newcommand{\gsim}{\mathrel{\hbox{\rlap{\lower.55ex \hbox{$\sim$}} \kern-.3em \raise.4ex \hbox{$>$}}}}

\newcommand{\ff}{g}
\newcommand{\DD}{D}
\newcommand{\smic}{\sigma^2_{\!\mathrm{mc}}(t)}
\newcommand{\mic}{\mathrm{mc}}

\newcommand{\hr}{{\cal H}}
\newcommand{\ord}{{\cal O}}

\newcommand{\diffa}{B}

\newcommand{\tr}{\mbox{Tr}}
\newcommand{\da}{\Delta_{\! A}}
\newcommand{\ds}{\Delta_{S}}

\newcommand{\dda}{\delta \! A}

\newcommand{\dds}{\delta S}

\begin{document}

\title{Generalization of von Neumann's Approach to Thermalization}

\author{Peter Reimann}
\affiliation{Fakult\"at f\"ur Physik, Universit\"at Bielefeld, 
33615 Bielefeld, Germany}

\begin{abstract}
Thermalization of isolated many-body 
systems is demonstrated by generalizing an approach originally due to 
von Neumann:
For arbitrary initial states with a macroscopically well-defined
energy, quantum mechanical expectation values 
become indistinguishable from the corresponding 
microcanonical expectation values for the overwhelming 
majority of all sufficiently late times.
As in von Neumann's work, 
the eigenvectors of the Hamiltonian and of the considered 
observable are required to not exhibit any specially tailored 
(untypical) orientation relative to each other.
But all of von Neumann's further assumptions about the 
admitted observables are abandoned.
\end{abstract}


\maketitle

The universal and irreversible tendency of 
nonequilibrium states towards thermal equilibrium 
is an everyday experience in the macroscopic 
world, but in spite of more than a century of 
theoretical efforts, it has still not been satisfactorily
reconciled with the basic laws of physics,
which govern the microscopic world, 
and which are fundamentally reversible \cite{skl93}.
The first quantum mechanical exploration 
of this problem is due to von Neumann \cite{neu29},
was unfortunately misunderstood for decades,
but has recently been rehabilitated 
in a very
enlightening commentary by Goldstein, Lebowitz,
Tumulka, and Zangh\`{\i} \cite{gol10a}.
A major remaining bottleneck of von Neumann's 
approach is his notion of ``macro-observer'' 
or ``macroscopic measurement'',
stipulating that all relevant
observables can be approximated by 
commuting Hermitian operators
with very high-dimensional common eigenspaces \cite{neu29}.
As an alternative, Goldstein et al. 
\cite{gol10a,gol10b} suggested to
consider ``macroscopic observables'' 
with the additional property 
(excluded in von Neumann's original treatment)
that one of those eigenspaces is 
overwhelmingly large compared to all the others.
In our present work,
all such restrictions with respect to the considered
observables are abandoned.

As in Refs. \cite{neu29,gol10a,gol10b}, we consider 
an isolated many-body system,
whose energy $E$ is known up to an
uncertainty $\delta E$, which is
small on the macroscopic but
large on the microscopic scale.
The system is modeled by a
Hamiltonian $H$ with 
eigenvalues
$E_n$ and eigenvectors $|n\rangle$,
$n\in{\mathbb N}$.
System states (pure or mixed) are described 
by density operators $\rho$, 
evolving in time according to 
the usual Liouville-vonNeumann equation
$\dot\rho(t)=i[\rho(t),H]/\hbar$.
Observables are modeled by Hermitian
operators $A$ with expectation values
$\langle A\rangle_\rho:=\tr\{\rho A\}$.
The preset energy interval $[E,E+\delta E]$ 
defines an energy shell, namely
the Hilbert space $\hr$ spanned by all 
$|n\rangle$ with $E_n\in [E,E+\delta E]$.
Without loss of generality, we assume that
the corresponding labels  are $n=1,2,\ldots,\DD $.
For a macroscopic system with, say, 
$f \approx 10^{23}$ degrees of freedom, the 
dimensionality $\DD $ of $\hr$ is exponentially 
large in $f$
\cite{gol10a}, 
symbolically indicated as
\begin{equation}
\DD  \approx 10^{\ord(f)} \ .
\label{1}
\end{equation}

By definition, the probability to encounter a 
system energy outside $[E,E+\delta E]$ is
negligibly small, and is henceforth idealized
as being strictly zero.
As a consequence, the diagonal matrix 
elements (``level populations'')
$\rho_{nn}:=\langle n|\rho|n\rangle$ 
vanish for all $n>\DD $, implying 
with Cauchy-Schwarz's inequality
that $\rho_{mn}=0$ if $m>\DD $ or 
$n>\DD $.
Denoting by $P$ the projector onto 
$\hr$, the projection (or restriction)
of $A$ onto $\hr$ takes the form
$\tilde A:=PAP$ and analogously 
$\tilde H:=PHP$ etc.
(note that $\tilde \rho=\rho$).
It readily follows that  
$\tr\{\rho A\}=\tr\{\rho \tilde A\}$
and that $\tilde H$ yields the same
time evolution of $\rho(t)$ as $H$.
Hence we can and will restrict ourselves
to the energy shell $\hr$ from 
now on, but, for convenience, 
omit the tilde symbols.
Accordingly, $P$ becomes the identity 
operator on $\hr$ and the microcanonical 
density operator follows as
$\rho_{\mic}:=P/\DD $ with expectation values 
$\langle A\rangle_{\mic}:=\tr\{\rho_{\mic}A\}$.

The problem of thermalization is to 
show that $\langle A\rangle_{\rho(t)}$
evolves towards $\langle A\rangle_{\mic}$
for arbitrary (possibly far from equilibrium) 
initial conditions $\rho(0):\hr\to\hr$.
It is well known that this is impossible without
additional assumptions on $H$ and $A$.
With respect to $H$, we adopt von Neumann's
assumption \cite{neu29} that the energy
differences $E_m-E_n$ are finite and mutually 
different for all pairs $m\not =n$.
Excluding nongeneric cases with
additional conserved quantities (besides $H$),
e.g. due to (perfect) symmetries or noninteracting subsystems,
the validity of this assumption is by now commonly accepted
\cite{per84,tas98,sre99,gol06,rei08,lin09,rei10,gol10b}.
Moreover, one expects that even considerably weaker assumptions 
will do \cite{sho11,sho12,rei12a,rei12b}.

Denoting by $a_{max}$ and $a_{min}$ the largest
and smallest among the ${\DD }$ eigenvalues of $A$,
the range of 
$A$ is defined as
$\da:=a_{max}-a_{min}$.
Furthermore, in any real  (or numerical) experiment, 
$\langle A\rangle_{\rho}$ can be determined only
with some finite accuracy $\dda$.
In practice, we thus can focus on measurements 
which yield at most, say, 20 relevant digits, i.e.
\begin{equation}
\dda \geq \da \,10^{-20} \  .
\label{2}
\end{equation}

The eigenvectors 
of $H$ and of $A$
are related by some unitary
basis transformation $U$. 
A key point of von Neumann's approach
is the assumption that these two
eigenbases do not exhibit any ``special
orientation'' relative to each other \cite{neu29,gol10c},
i.e., the actual $U$ is ``typical'' \cite{gol10a}
among all possible unitary transformations 
$U:\hr\to\hr$ in the following sense:
If a certain property can be shown
to hold for the vast majority of $U$'s 
(uniformly distributed according to the 
Haar measure \cite{neu29,gol10a,gol10b,gol10c}),
then this property is supposed to 
hold for the actual $U$ as well, unless 
there are special 
reasons to the contrary.
Denoting by $\mu_U(X)$ the fraction 
(normalized measure) of all $U$'s exhibiting 
a certain property $X$,
a $\mu_U(X)$ value
close to unity (zero) is thus assumed to 
generically imply (exclude) property
$X$ for the actual system.
While a more rigorous justification is clearly
very difficult, intuitively such a ``typicality''
argument is very convincing:
If we imagine $A$ as fixed 
and $H$ as arising by randomly sampling
its eigenvectors via $U$
\cite{fie55,gol10b,gol10c},
the argument is essentially tantamount to
the common lore of random matrix theory 
\cite{gol10a}, which is well known to be extremely
successful in practice \cite{bro81}.
In particular, $\mu_U(X)$ may be 
{\em formally} viewed as the probability 
of observing property $X$ for a randomly 
sampled $H$ (or $U$), 
however keeping in mind 
-- exactly as in random matrix theory --
that there is 
{\em no random sampling procedure 
in the real physical problem under 
consideration} \cite{gol10a,gol10b}.
In passing we note that von Neumann actually
adopted the complementary viewpoint
of considering $H$ as fixed while varying 
the eigenvectors of $A$ \cite{neu29}.

By exploiting the above mentioned common
assumptions about 
%the energy differences $E_m-E_n$, 
the energy eigenvalues $E_n$
\cite{neu29,per84,tas98,sre99,gol06,rei08,lin09,rei10,gol10b},
one can infer
\cite{per84,sre99,tas98,rei08,lin09,rei10,sho11}
(see also \cite{sup})
that the quantity
\begin{equation}
\sigma^2(t):=
 \left[ \langle A\rangle_{\rho(t)}  - \langle A\rangle_{\bar\rho} \right]^2 
\label{3}
\end{equation}
satisfies the relations
\begin{equation}
\overline{ \sigma^2(t) }
= \sum_{m\not=n}^{\DD}  |\rho_{mn}{(0)}|^2 \, |A_{mn}|^2
\leq \max_{m\not=n}  |A_{mn}|^2 \ ,
\label{4}
\end{equation}
where 
$A_{mn}:=\langle m |A| n \rangle$,
$\rho_{mn}{(0)}:=\langle m |\rho(0)| n \rangle$,
and the overbar indicates an average over all 
times $t\geq 0$.
In particular, $\bar{\rho}:=\overline{\rho(t)}$
is an auxiliary density operator with matrix elements
$\bar{\rho}_{mn} = \delta_{mn} \rho_{nn}{(0)}$,
sometimes named diagonal or 
generalized Gibbs ensemble \cite{rig08}.
The so-called eigenstate thermalization hypothesis
(ETH) {\em conjectures} that for a many-body system
with $f\gg 1$ degrees of freedom, 
typical off-diagonal elements
$A_{mn}$ in (\ref{4}) are exponentially small in $f$
\cite{deu91,sre94,sre99,rig08,gol11,rig12}.
Within our present generalization of von Neumann's
approach, we can actually {\em prove} that even 
their maximum on the right-hand side in (\ref{4}) is 
typically so small that
\begin{equation}
\mu_U\left( \overline{ \sigma^2(t) } \geq\epsilon\right)
\leq 4\, \exp\left\{-
\frac{\epsilon \DD }{18\pi^3\da^2}+2\ln \DD \right\} 
\label{5}
\end{equation}
for any $\epsilon>0$.
Besides (\ref{4}), the key ingredient in deriving this 
result is Levy's lemma (see \cite{pop06,pop06a} and 
further references therein),  stating that
\begin{equation}
\mbox{Prob}\Bigl( \left| \ff (\phi) - \langle \ff \rangle \right| 
\geq\epsilon\Bigr)
\leq 2\, \exp\left\{-
\frac{\epsilon^2 (d+1)}{9\pi^3\eta^2}\right\}
\label{6}
\end{equation}
for randomly and uniformly distributed
points $\phi$ on the $d$-dimensional 
unit sphere ${\mathbb S}^d\subset {\mathbb R}^{d+1}$ 
and any Lipschitz continuous function 
$\ff :{\mathbb S}^d\to{\mathbb R}$ 
with Lipschitz constant $\eta$
and mean value $\langle \ff \rangle$.
Furthermore, any normalized $|\phi\rangle\in\hr$ 
of the form $\sum_{n=1}^{\DD } c_n|n\rangle$
can be represented (via the real and imaginary
parts of the $c_n$'s) as a point $\phi$
on the $(2\DD-1)$-dimensional unit sphere.
Finally, one can show \cite{pop06a}
that $\ff (\phi):=\langle \phi|A|\phi\rangle$ 
is Lipschitz continuous 
with $\eta=\da$ and 
$\langle \ff \rangle=\langle A\rangle_\mic$.
Observing that randomizing $\phi$ is 
equivalent to randomizing $U$, 
we thus obtain
\begin{equation}
\mu_U\Bigl( \left|
\langle \phi|A|\phi\rangle
-\langle A\rangle_\mic \right| 
\geq\epsilon\Bigr)
\leq 2\, \exp\left\{-
\frac{2\, \epsilon^2 \DD }{9\pi^3 \da^2}\right\} \ .
\label{7}
\end{equation}
The remaining task is to connect this result for 
$\langle \phi|A|\phi\rangle$  with the maximal $|A_{mn}|$ in (\ref{4}).
The details are rather straightforward but tedious and thus
provided as supplemental material in \cite{sup}.
As an aside, it follows that von Neumann's main
technical achievement (Appendix of \cite{neu29}),
as well as its further improvement by 
Pauli and Fierz \cite{pau37},
is in fact quite closely related to Levy's lemma
(see also \cite{sup}).

Equation (\ref{5}) represents the first main result
of our Letter.
By choosing, e.g., 
$\epsilon=\DD ^{-1/2}\dda^2$ 
in (\ref{5}), it 
follows with (\ref{1}) and (\ref{2}) that 
the time-averaged variance from (\ref{3})
remains extremely much smaller than 
$\dda^2$ for ``almost all'' $U$:
The fraction of the exceptional $U$'s 
is an unimaginably small number of the 
order of $10^{-x}$ with $x\approx 10^{\ord(f)}$, 
$f\approx 10^{23}$.
Furthermore, the mere existence of the infinite time 
average in (\ref{4}) implies that a similar estimate 
must also apply to averages of $\sigma^2(t)$ 
over finite time intervals $[0,T]$ with sufficiently 
large $T$ \cite{tas98,gol10a,sho12,rei12a}.
Finally, the smallness of the latter time 
average implies (obviously or by Markov's 
inequality) that the averaged quantity (\ref{3})
itself must be exceedingly small for most times 
$t \in [0,T]$ \cite{tas98,lin09,gol10a,rei12a}.
For example, for our above choice $\epsilon=\DD ^{-1/2}\dda^2$
and sufficiently large $T$,
all the ``bad times'' $t\in[0,T]$ with 
$|\langle A\rangle_{\rho(t)} 
- \langle A\rangle_{\bar\rho}|\geq \dda$ 
add up to a set, whose Lebesgue measure
is smaller by (at least) a factor of the order 
$\DD ^{1/4}\approx 10^{\ord(f)}$
(cf. (\ref{1})) than the measure of all 
$t\in[0,T]$.
Altogether, we thus can conclude that for the overwhelming 
majority of $U$'s, the difference
$\langle A\rangle_{\rho(t)} - \langle A\rangle_{\bar\rho}$ 
remains below the resolution limit $\dda$ for the vast 
majority of times $t$ 
contained in any sufficiently large time 
interval $[0,T]$.
The same conclusion carries over to our actual
Hamiltonian $H$ and observable $A$, given
their eigenbases are related by a ``typical'' 
transformation $U$ as discussed above.
To establish quantitative bounds for $T$
is a subject of considerable current
interest
\cite{sre99,sho12,gol13,mon13,mal14,gol15},
but goes beyond our present scope.

The salient point is that (\ref{5}) holds 
independently of the initial condition $\rho(0)$. 
Once a pair $H$, $A$ with a typical $U$
is given, the above implications of (\ref{5}) 
thus apply to {\em any} $\rho(0)$:
No matter how far from equilibrium 
the system starts out, for almost all sufficiently 
late times it behaves practically
as if it were in the state $\bar\rho$.
Such an apparent convergence towards a 
steady state has been denoted as 
{\em equilibration}, e.g., in Refs.
\cite{rei08,lin09,rei10,sho11,sho12,rei12a,rei12b}.

To demonstrate {\em thermalization}, we still 
have to show that the difference between
$\langle A\rangle_{\bar\rho}$ and 
$\langle A\rangle_{\mic}$ 
is negligibly small.
Recalling the definitions of these two expectation
values, one readily sees that
\begin{equation}
\diffa:=\langle A\rangle_{\bar\rho} -\langle A\rangle_{\mic}
=
\sum_{n=1}^{\DD}  {\rho}_{nn}{(0)}\,
[A_{nn} - \langle A\rangle_{\mic}]
\label{8}
\end{equation}
and hence that
\begin{equation}
|\diffa|
\leq
\max_{n}\left| A_{nn} - \langle A\rangle_{\mic} \right| \ .
\label{9}
\end{equation}
Similarly as above Eq. (\ref{5}), one part (actually the better
known part) of ETH consists in the {\em conjecture} that
typical differences $A_{nn} - \langle A\rangle_{\mic}$ 
are exponentially small in $f$ \cite{deu91,sre94,sre99,rig08,gol11,rig12}.
Within our present framework, we can {\em prove}
that even their maximum
in (\ref{9})
is typically so small that
\begin{equation}
\mu_U\left(
| \diffa | 
\geq \epsilon\right)
\leq 2\, \exp\left\{-\frac{2}{9\pi^3} \frac{\epsilon^2 \DD }{\da^2}+\ln \DD \right\} 
\label{10}
\end{equation}
for any $\epsilon>0$.
This represents our second main result, whose
derivation from (\ref{7}) is quite obvious and
is provided in full detail as supplemental material in \cite{sup}.
Once again, it is crucial to note that
(\ref{10}) is independent of $\bar\rho$
(and thus of $\rho(0)$):
Given a pair $H$, $A$ with a typical $U$,
it follows from (\ref{8}) and (\ref{10})
that the difference $\langle A\rangle_{\bar\rho}-\langle A\rangle_{\mic}$ 
remains way below the resolution limit $\dda$
for {\em any} $\bar\rho$ (or $\rho(0)$).
Finally, upon considering $A$ as fixed and $H$ as a 
random matrix (see above) we can conclude that 
von Neumann's approach \cite{neu29} in fact anticipates
the 
verification of ETH 
from Ref. \cite{deu91} within a random matrix theoretical
framework, see also 
\cite{fie55,gol10a}.

One readily sees that the measure of all $U$'s 
which give simultaneously rise to both equilibration, 
as discussed in the paragraph below (\ref{7}),
{\em and} negligibly small $\diffa$ values according to 
(\ref{10}) is still extremely close to unity.
Hence thermalization follows for any given
pair $H$, $A$ with a generic relative orientation
of the eigenbases, 
no matter how the initial condition 
$\rho(0)$ is chosen.
Along the same lines, one can infer
the simultaneous thermalization of
several (not necessarily commuting)
observables \cite{rei07,sho11},
as long as their number remains 
``reasonable'' (e.g., smaller than $\DD $).

Similarly as for $U$, let us now denote by $V$ 
the unitary basis transformation between the 
eigenvectors of the density operator $\rho(0)$ 
and those of $H$.
Likewise, $\mu_V(X)$ now represents
the fraction (normalized measure) of all unitary 
transformations $V:\hr\to\hr$ which 
exhibit a certain property $X$.
Furthermore, the usual von Neumann entropy 
%of any (mixed or pure) state $\rho$ 
is defined as $S[\rho]:=-k_B\tr\{\rho\ln\rho\}$
and satisfies $0\leq S[\rho]\leq S[\rho_{\mic}]=k_B \ln \DD $.
Hence the entropy range is $\ds=k_B \ln \DD $ and, 
similarly as in (\ref{2}), experimentally resolvable 
entropy differences $\dds$ can be assumed to 
satisfy $\dds\geq q\ds = q\, k_B \ln \DD $
for some small but still ``reasonable'' 
$q$ value.
It follows that $\rho(0)$ entails a $\bar\rho$
with the properties that $S[\rho_{\mic}]-S[\bar\rho]\geq 0$
and, as demonstrated in detail in the 
supplemental material \cite{sup},
\begin{equation}
\mu_V\left( S[\rho_{\mic}]-S[\bar\rho]\geq s \right)
\leq k_B/s
\label{11}
\end{equation}
for any $s>0$.
This is our third main result.
By choosing $s=\dds$ and recalling that
$\dds\geq q \,k_B \!\ln \DD $ (see above),
it implies with (\ref{1}) that the entropy of the 
diagonal ensemble $\bar\rho$, towards which 
the ``true'' $\rho(t)$ seems to equilibrate, 
differs from the microcanonical entropy only 
by an unmeasurably small amount for a generic
$\rho(0)$, i.e. one without a specially tailored 
orientation of its eigenbasis relative to that of $H$.
We remark that  already von Neumann demonstrated
a somewhat similar, so-called $H$ theorem \cite{neu29}, 
however, for a differently defined entropy, 
whose physical relevance has been questioned, 
e.g., in Ref. \cite{gol10a}.
Further related but different results about entropies 
of diagonal ensembles are also due to \cite{pol11}.

As shown in Refs. \cite{rei08,lin09,rei10,sup}, an alternative 
upper bound for the left-hand side of (\ref{4}) is 
given by $(\da^2/4) \,\tr\{{\bar\rho}^2\}$.
For the latter factor, $\tr\{{\bar\rho}^2\}$, a similar 
relation as in (\ref{11}) is derived 
in the supplemental material \cite{sup},
yielding
\begin{equation}
\mu_V \left( \overline{\sigma^2(t)} \geq\epsilon\right)\leq 
\da^2/(2\epsilon \DD )
\label{12}
\end{equation}
for any $\epsilon>0$.
By analogous arguments as 
in the discussion of (5)
in the paragraph below (\ref{7}),
this amounts to an alternative 
demonstration of equilibration 
\cite{tas98,rei08,lin09,rei10}.
But in contrast to (\ref{5}), which applies
to arbitrary $\rho(0)$, provided the relative
eigenbasis orientation of $H$ and $A$ 
is generic,
the present findings now apply 
to arbitrary 
observables $A$,
provided the eigenbases of $H$ and $\rho(0)$ 
are in a generic constellation.

Finally, let us denote by $W$
the unitary basis transformation
between the eigenvectors of the
density operator $\rho(0)$ and 
those of $A$,
and consider
\begin{equation}
\smic:=
 \left[ \langle A\rangle_{\rho(t)}  - \langle A\rangle_{\mic} \right]^2 \ .
\label{13}
\end{equation}
Similarly as before, we now can show (see 
supplemental material \cite{sup}) that
\begin{equation}
\mu_W\left( \smic \geq\epsilon\right)
\leq \da^2/(\epsilon \DD )
\label{14}
\end{equation}
for arbitrary $t$ and $\epsilon>0$,
and that
\begin{equation}
\mu_W\left( \frac{1}{t_2-t_1}
\int_{t_1}^{t_2} \smic \,\mbox{d}t
\geq\epsilon\right)
\leq \da^2/(\epsilon \DD )
\label{15}
\end{equation}
for arbitrary $t_1<t_2$.
Note that while (\ref{14}) is a $t$ independent
upper bound for the {\em measure} of all $W$'s 
with 
%the property that 
$\smic\geq\epsilon$, 
this does not imply that the {\em set} of those 
$W$'s is $t$-independent.
An analogous caveat applies to
(\ref{15}).
Yet another crucial point
is that (\ref{14}) and (\ref{15}) 
are valid for completely arbitrary (even time-dependent)
Hamiltonians $H:\hr\to\hr$ 
\cite{boc58,boc59,sup}.

A first remarkable implication of (\ref{14}) and (\ref{15}) 
follows by considering $A$ as ``given'' (arbitrary but fixed):
Namely, ``most'' \cite{f1}
$\rho(0)$ then yield practically 
the same expectation value $\langle A\rangle_{\rho(t)}$
as  $\rho_{\mic}$ for any arbitrary but fixed 
time point $t$,
but also for practically all times $t$ 
within an arbitrary but fixed 
time interval $[t_1,t_2]$ (see paragraph below (\ref{7})).
Put differently, nonequilibrium 
expectation values 
are ``untypical'' (even for $t=0$), 
they require very special 
orientations $W$ of the eigenbasis 
of $\rho(0)$ relative to that of $A$.
In particular, for pure states 
$\rho(0)=|\psi \rangle\langle \psi|$
we recover the quintessence of so-called
canonical typicality and related phenomena 
\cite{pop06,gol06b,rei07,bar09,sug12}
(see also \cite{boc59,pec84}).

Conversely, when considering $\rho(0)$ as
``given'', it follows from (\ref{14}) and (\ref{15})
that ``most'' \cite{f1} measurement devices $A$ 
cannot distinguish $\rho(t)$ from 
$\rho_{\mic}$ at any arbitrary but fixed time 
point $t$, or for practically all times $t$ within 
an arbitrary but fixed time interval $[t_1,t_2]$.
This is the viewpoint adopted, e.g., in Refs.
\cite{far57,boc58},
but now formulated within our present
generalization of von Neumann's 
original approach (see also \cite{mal14}).

The fact that a nonequilibrium 
value of $\langle A\rangle_{\rho(0)}$
requires an untypical pair $\rho(0)$, $A$
implies \cite{gol10a} that conclusions regarding 
%non-equilibrium initial states and their
thermalization can 
be drawn only from results concerning {\em all} 
orientations of $\rho(0)$ relative to $A$,
as it is the case in von Neumann's approach
(see below (\ref{7}) and (\ref{10})),
but not from results concerning
{\em most} orientations, as in 
the above generalization 
(\ref{14}) and (\ref{15}) of the approach 
from Refs. \cite{far57,boc58,boc59,pec84}.
In other words, it is not right to
say that von Neumann's approach is 
inadequate to investigate thermalization
since the same applies to the approach  
from Refs. \cite{far57,boc58,boc59,pec84}.
Rather, the two approaches are fundamentally 
different:
One requires a generic eigenbasis 
constellation of $H$ and $A$ but admits 
any $\rho(0)$,
the other requires a generic eigenbasis 
constellation of $\rho(0)$ and $A$ but 
admits any $H$.

In conclusion, von Neumann's demonstration of 
thermalization for isolated many-body systems 
has been generalized to arbitrary observables.
The remaining prerequisites for 
thermalization are thus rather weak, namely
a Hamiltonian 
with generic eigenvalues 
$E_n$ and a generic orientation of its 
eigenvectors relative to those of $A$, 
while the initial state $\rho(0)$ may still be chosen 
arbitrarily (mixed or pure, far from equilibrium or not).
The first requirement (regarding $E_n$) 
is by now well established 
\cite{per84,tas98,sre99,gol06,rei08,lin09,rei10,gol10b},
and further generalizations like in Refs.
\cite{sho11,sho12,rei12a,rei12b}
seem possible.
With the second requirement, 
von Neumann essentially anticipated the 
foundation of random matrix theory \cite{gol10a},
which is very difficult to justify rigorously,
but is extremely successful in practice,
and can be corroborated by various 
intuitively convincing arguments \cite{bro81}.
For instance:
Since our mind is used to thinking about the 
physical world in terms of  individual ``particles'', 
we mostly come up with single-particle observables $A$
or sums thereof (kinetic energy (temperature), 
density, pressure, magnetization, etc.),
whose eigenvectors are thus
single-particle product states.
In contrast, a generic Hamiltonian $H$ includes
particle-particle interactions, giving rise
to a ``completely different'' eigenbasis
without any ``special relation'' to that 
of $A$ \cite{per84}.

While von Neumann had in mind a preset 
$H$ and a varying (or typical) eigenbasis of 
$A$ \cite{neu29},
the mathematically equivalent but physically 
opposite viewpoint (fixed $A$, varying $H$)
was emphasized, e.g., in Refs. 
\cite{fie55,gol10a,gol10b,gol10c}.
Here, both views have been merged and
significantly generalized by treating 
all three operators $A$, $H$, and $\rho(0)$ 
on an equal footing:
After selecting two of them and assuming 
they exhibit a typical eigenbasis constellation, 
we were able to draw conclusions which are then 
entirely independent of the third one.
Along these lines, we established the 
general new results (\ref{11}) and (\ref{12}) 
concerning the generic long-time behavior 
(equilibration) for expectation values of
arbitrary (even untypical) observables 
and the entropy of the concomitant
equilibrium states.
Furthermore, our findings (\ref{14}), (\ref{15}) 
significantly generalize previously 
known typicality results for arbitrary 
(even time-dependent) Hamiltonians. 
As a by-product we thus
obtained a unifying framework for several 
key aspects 
of thermalization,
such as the validation of the
eigenstate thermalization hypothesis 
\cite{sre94,sre99,rig08,rig12}
by means of random
matrix theory \cite{deu91},
recent explorations of ``equilibration'' 
\cite{rei08,lin09,rei10,sho11,sho12,rei12a,rei12b}
and ``canonical typicality'' 
\cite{pop06,gol06b,rei07,bar09,sug12},
the long-lasting misjudgment of von Neumann's work
\cite{far57,boc58,boc59,pec84},
and its rehabilitation in Ref. \cite{gol10a}.
\begin{center}
\vspace{5mm}
---------------------------
\vspace{5mm}
\end{center}
This work was supported by DFG-Grant RE1344/7-1.


\begin{thebibliography}{11}
\bibitem{f1}
In principle, ``most'' is meant with 
respect to eigenbasis variations via $W$,
but 
different spectra can also be readily included.
\end{thebibliography}
\end{document}